\documentclass{kapproc} 

\usepackage{procps} 
\usepackage[latin1]{inputenc}

\usepackage[dvips]{graphicx}

\upperandlowercase

\kluwerbib

\begin{document}

\articletitle[Internal Kinematics of LCBGs]
{Internal Kinematics of Luminous Compact Blue Galaxies}
\author{Matthew A. Bershady,\altaffilmark{1} 
M. Vils,\altaffilmark{1} C. Hoyos,\altaffilmark{2}
R. Guzm\'an,\altaffilmark{3} D. C. Koo\altaffilmark{4}}
\affil{\altaffilmark{1}U. Wisconsin, \ \altaffilmark{2}U. Autonoma de Madrid, \
\altaffilmark{3}U. Florida, \ \altaffilmark{4}U. California, Santa-Cruz}
\begin{abstract}
We describe the dynamical properties which may be inferred from
HST/STIS spectroscopic observations of luminous compact blue galaxies
(LCBGs) between $0.1<z<0.7$. While the sample is homogeneous in blue
rest-frame color, small size and line-width, and high
surface-brightness, their detailed morphology is eclectic. Here we 
determine the amplitude of rotation versus random,
or disturbed motions of the ionized gas.  This information affirms the
accuracy of dynamical mass and M/L estimates from Keck integrated
line-widths, and hence also the predictions of the photometric fading
of these unusual galaxies. The resolved kinematics indicates this
small subset of LCBGs are dynamically hot, and unlikely to be embedded
in disk systems.
\end{abstract}

\section{Introduction}

The evolution of LCBGs is a matter of debate. These galaxies are
unusual in their blue colors, small sizes and line-widths, yet large
luminosities.  We have suggested that at least a subset of these
sources are the progenitors of dEs such as NGC 205 (Koo et al. 1995,
Guzman et al. 1998), while others counter these are bulges in
formation (Hammer et al. 2001; Barton \& van Zee 2001). Surveys at
intermediate redshift are not uniformly defined, and each contains
heterogeneous samples -- objects span a range in size, color,
luminosity, surface-brightness, and image concentration.  The broad
``LCBG'' class contributes as much as 45\% of the comoving SFR between
$0.4<z<1$ (Guzman et al. 1997); the proposed dE progenitors are a
fraction of this class. Here we focus on an extreme LCBG sub-class
that are among the smallest, bluest and highest surface-brightness
(Koo et al. 1995): M$_B$$\sim$$-21$ (H$_0$=70 km/s/Mpc, $\Omega$=1,
$\Omega_\Lambda$=0.7), rest-frame B-V$\sim$0.25, half-light radii of
$R_e$$\sim$2 kpc, mean surface-brightness within $R_e$ of $\sim$19
mag/arcsec$^2$ (rest-frame B band), and integrated line-widths of
$\sigma_{gas}$$\sim$65 km/s. Many of these are good candidates for dE
progenitors. If so, their internal kinematics should reveal they are
dynamically hot, while deep imaging should show they lack outer disks.

\section[Hot or Cold?]
{STIS Spectra: Are LCBGs Dynamically Hot or Cold?}

We have derived ionized-gas position--velocity and
position--line-width diagrams from STIS long-slit measurements along
what appears to be the photometric major axes of 6 LCBGs between
$0.2<z<0.7$, and one other source at $z\sim0.1$ which is 2-3 mag lower
luminosity than the others. One example is given in Figure 1. With 0.2
arcsec slits, STIS delivers instrumental resolutions ($\sigma$) of
13-19 km s$^{-1}$. Line-emission is not always centered on the
continuum (Hoyos et al. 2004); the continuum centroid is adopted as
the kinematic center.

We find Keck HIRES integrated line-widths (Koo et al 1995) agree in
the mean with the resolved velocity dispersions from STIS
spectroscopy: Integrated dispersions are not due to large-scale, bulk,
motion. This secures our previous dynamical estimates of M/L and their
use as constraints on photometric fading (e.g., Guzman et
al. 1998). Only the low-$L$, low-$z$ system shows clear rotation 
and substantially different integrated versus resolved line-widths.

\begin{figure}[ht]
\vskip -0.45in
\rotatebox{-90}{
\includegraphics[width=4in]{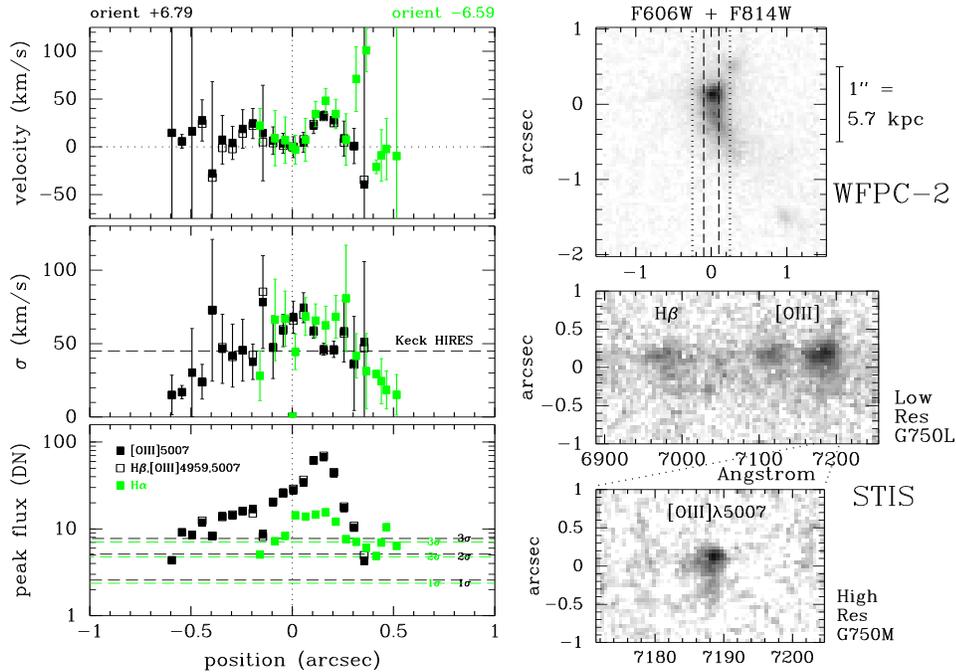}}
\vskip -0.05in
\caption{Morphology and kinematics of LCBG 313088 at $z=0.44$.  Left:
HST/WFPC-2 image showing distorted, tail-like source morphology and
mean STIS slit positions (0.2 and 0.5 arcsec).  STIS spectra for low
and high-resolution gratings are at bottom. Right: Position vs
velocity, line-width, and line-flux for two sets of high-resolution
data taken at two position angles varying by
$\sim13^\circ$. (H$\alpha$ spectrum not shown). Spectral data
consistently show extended line-emission with little velocity
gradient, no evidence for rotation, and dispersions that agree in the
mean with Keck HIRES integrated measurements (dashed line, middle
panel).}
\end{figure}

\begin{figure}[ht]
\vskip -0.9in
\includegraphics[width=\textwidth]{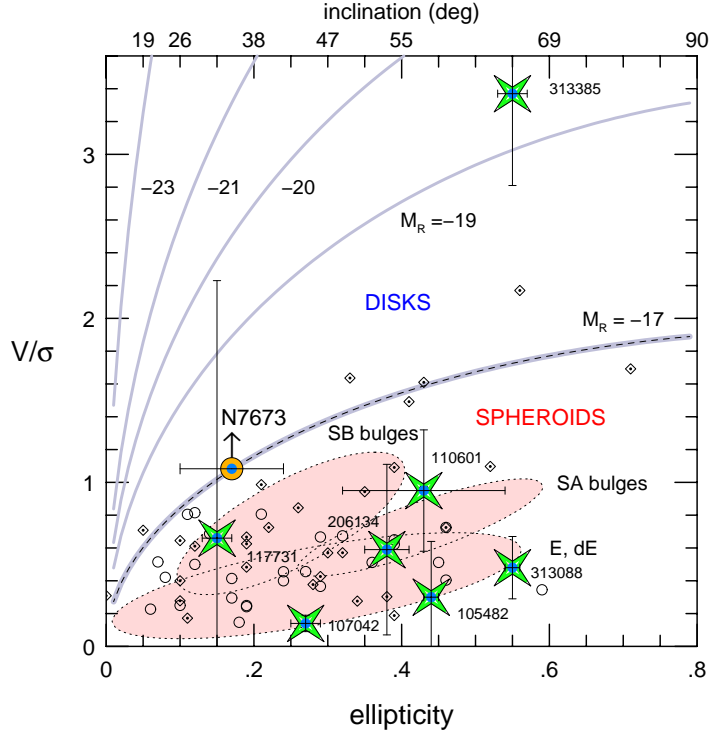}
\vskip -1.6in
\caption{V/$\sigma$ versus ellipticity or inclination for LCBGs
(dotted stars), NGC 7673 (circle/lower limit) and early-types galaxies
from Simien \& Prugniel (circles, dotted diamonds). Lines represent
trajectories as a function of inclination for disk-galaxies of
different luminosity assuming they lie on the Tully-Fisher relation and
$\sigma_{gas} = 25$ km s$^{-1}$ (Andersen et al. 2005).
Shaded regions are adopted from Kormendy \& Kennicutt (2004).
}
\vskip -0.1in
\end{figure}

The lack of rotation coupled with their ellipticity squarely places
these sources in the ``spheroidal'' region of the V/$\sigma$ -
ellipticity plane, illustrated in Figure 2.  For fair comparison to
local samples, ellipticities are measured at the half-light radius
near rest-frame V-band from HST images; rotation velocities are set to
half the difference between minimum and maximum velocities; and
$\sigma$ is the observed central velocity dispersion. While there is a
range of observed V/$\sigma$, LCBGs lie well below the region
inhabited by disk systems, with values comparable to local dEs and
other spheroidals, particularly if recent observations of dEs with
larger rotational components are considered (Pedraz et al. 2002, van
Zee et al. 2004): {\bf LCBGs are dynamically HOT}.

\section[Scenarios]
{Progenitors and Descendants}

Do we know that we aren't just sampling a bulge, or a face-on nuclear
star-burst? NGC 7673 has been suggested by Homeier et al. (2002) as a
nearby example. Indeed this source has the right color,
surface-brightness, size, luminosity and integrated line-width, and
has a faint, extended outer disk. However, Homeier \& Gallagher's
(1999) H$\alpha$ velocity map shows clear evidence for rotation in the
inner, star-burst region. Our re-analysis confirms this: We would see
similar structure {\it if it existed} in our STIS
spectra. Independently, deep CFHT imaging reveals no strong evidence
for extended, normal disks around the types of (and specific) sources
presented here (Barton et al. 2005).

In summary, the preponderance of evidence is against bulge formation
in disk systems and in favor of a dE-like descendant scenario {\it for
all of the specific sample presented here} with M$_B$$<$$-20$ (6 out
of 7). However, the disturbed morphology and kinematics makes clean
interpretations difficult. What is the gas really telling us about
dynamics? Are these systems in dynamical equilibrium? While their
morphology and resolved kinematics would argue otherwise, the
agreement between integrated velocity dispersions and resolved
profiles indicates the systems cannot be too far out of
equilibrium. Stellar velocity and dispersion profiles would provide a
much clearer dynamical picture.

Finally we comment on issues raised at the conference about
environment: Is NGC 205 a good example of a faded, LCBG descendant?
If so, where are the M31-like neighbors? Are there field dEs?  If
dSphs are the low-mass cousins of dEs, then the presence of isolated
dSphs such as Tucana, Cetus, and the recently discovered Apples 1
(Pasquali et al. 2004) should give us pause about accepting assertions
that dEs do not exist outside of rich environments or far from massive
galaxies. The space density of LCBGs presented here is
1.25$\pm$0.15$\times10^{-5}$ Mpc$^{-3}$ for M$_B<-20$ between
$0.3<z<0.7$.  Even allowing a factor of $\sim20$ higher relic density
(given the time interval in this redshift slice and assuming the LCBG
phase is a few$\times$10$^8$ yr)
to find even one descendant requires an all-sky local survey volume
reaching out to $\sim$10 Mpc. At this distance the half-light radius
of NGC 205 is $\sim$12 arcsec. Are our local surveys this complete?

\end{document}